\documentclass[twocolumn,showpacs,preprintnumbers,aps,prl,10pt,nofootinbib]{revtex4}

\usepackage[english]{babel} 	
\usepackage{graphics} 
\usepackage[pdftex]{graphicx}
\usepackage{amssymb} 			
\usepackage{amsmath} 			
\usepackage{booktabs}
\usepackage{tabularx}
\usepackage{rotating}

\usepackage{float}
\usepackage{url}	
\usepackage{bm}

\graphicspath{{./figures/}}

\allowdisplaybreaks

\begin{document}


\title{A combined mean-field and three-body model tested on the
  $^{26}$O-nucleus}

\author{D. Hove$^{1}$, E. Garrido$^{2}$, P. Sarriguren$^{2}$,  D.V. Fedorov$^{1}$,
H.O.U. Fynbo$^{1}$, A.S. Jensen$^{1}$,  N.T. Zinner$^{1}$}

\affiliation{$^{1}$Department of Physics and Astronomy, Aarhus University, DK-8000 Aarhus C, Denmark}

\affiliation{$^{2}$Instituto de Estructura de la Materia, IEM-CSIC,
Serrano 123, E-28006 Madrid, Spain}

\date{\today}


\begin{abstract}
We combine few- and many-body degrees of freedom in a model applicable
to both bound and continuum states and adaptable to different
subfields of physics.  We formulate a self-consistent three-body model
for a core-nucleus surrounded by two valence nucleons.  We treat the
core in the mean-field approximation and use the same effective Skyrme
interaction between both core and valence nucleons.  We apply the
model to $^{26}$O where we reproduce the known experimental data as
well as phenomenological models with more parameters.  The decay of
the ground state is found to proceed directly into the continuum
without effect of the virtual sequential decay through the well
reproduced $d_{3/2}$-resonance of $^{25}$O.
\end{abstract}






\maketitle


\paragraph{Introduction}

Self-consistent mean-field calculations efficiently provide accurate
average properties for $N$-body systems \cite{ben75,nik11,dob16}.
Approximate methods have been developed to treat correlated systems
\cite{bed03,bed02,car16,bro88,bar13,pie04,ono92,fel97,hag10,suz90,tan78}.
All these methods are first of all aimed at describing bound states.
Resonances and their decay, and continuum states in general often are
addressed, but only with much more difficulty.  The challenges are
often referred to as problems in connection with open quantum systems
\cite{rot15}.  This concept is defined as quantum systems in
interaction with the environment via external fields or, more
appropriate in the present context, by coupling to continuum degrees
of freedom.

When $N$ is less than about $15$, complete correlated solutions can be
obtained with modern computers \cite{lei13,nav09,hag14,epe09,lie16}.
The solutions are obtained with few-body techniques, that is as
analytically or numerically accurate as necessary in the context. Thus, all necessary effects from
couplings to the continuum are in principle fully included.  These
few-body techniques are designed to operate for distances larger than
the radii of the constituent particles necessarily assumed to be
inert.  This is in contrast to the many-body methods designed to treat
approximately bound or at least quasi stable states at small
distances.

None of the existing methods treat both small-distance structure and
large-distance decay properties equally well.  The traditional
few-body cluster methods assume small-distance boundary conditions at
the surface of the constituents and provide corresponding
large-distance behavior.  The many-body methods provide detailed small
distance structures and therefore correct boundary conditions for
few-body calculations.

Combining these properties in one model would be extremely useful and
of general interest in all subfields of physics.  More specifically, a
number of interdisciplinary topical problems can be better understood
microscopically.  This applies in particular to the concept of
universality in connection with halo formation and decay \cite{fre12}
and the extreme of Efimov Physics \cite{bra06}.  Both phenomena appear
in nuclei and nuclear astrophysics, as well as in cold atomic and
molecular gases.

The purpose of this letter is to provide an overall framework to
combine the few- and many-body treatments of relative and intrinsic
motion of the constituent particles.  We shall use the hyperspherical
adiabatic Faddeev expansion method for the few-body part and the
mean-field approximation for the many-body part.  The effective
interactions in this combined model design is a challenge, because the
allowed many-body Hilbert space requires specially adjusted
interactions, which must be renormalized for use in the few-body
calculations.  The applied many-body effective interaction is in this
letter consistently incorporated in the few-body treatment, except for
two terms of minor importance where we indicate possible improvements.

The practical implementation is crucial to test applicability,
accuracy and efficiency.  We choose $^{26}$O, where the nucleons in
$^{24}$O-core and the additional two neutrons require different
treatment \cite{bro05}, and where the traditional methods are
inappropriate.  The $^{26}$O-nucleus is an ideal test case on the
neutron dripline with the double magic \cite{hof09} and spherical
$^{24}$O-core and the two valence neutrons \cite{cae13,koh13,kon16}.
A detailed phenomenological investigation has appeared in the course
of this work \cite{hag16}. We compare results and predict other
observables.

\paragraph{Theoretical formulation.}

We consider an $A+2$ nucleon system divided into a core with mass
number $A$ and two valence nucleons.  We assume the same two- and
three-body interactions, $V_{ij}$ and $V_{ijk}$, acting between all
the nucleons in core and valence space.  The general Hamiltonian can
then be written
\begin{align}
H=\sum_{i=1}^{A+2} T_i - T_{cm} + \sum_{i<j}^{A+2} V_{ij} + \sum_{i<j<k}^{A+2} V_{ijk}, 
\label{eq:fullHam}
\end{align} 
where $T_i$ and $T_{cm}$ are the kinetic energy operators for the $i'$th
nucleon and for the total $A+2$ system, respectively.  We reorganize
$H$ into terms related to core, $H_c$, and valence, $H_v$, particles,
i.e. explicitly 
\begin{eqnarray} \label{hamcv}
  H &=& H_c(\bm{r}_1,\cdots,\bm{r}_A)  + H_v(\bm{r}_{v_1},\bm{r}_{v_2})  \;,
  \\ \label{hamc}
  H_c  &=&   \sum_{i=1}^{A} T_i - T^{core}_{cm} +
  \sum_{i<j}^{A} V_{ij} + \sum_{i<j<k}^{A} V_{ijk} \;,  \\  \nonumber
   H_v &=& T^{core}_{cm} + T_{A+1} + T_{A+2} - T_{cm} + V_{A+1,A+2}  \\ \label{hamv}
   &+& \sum_{i=1}^{A} (V_{i,A+1} + V_{i,A+2})  + \sum_{i=1}^{A} V_{i,A+1,A+2}   \;, 
\end{eqnarray}
where the spin and space coordinates of the $i$'th core or valence
nucleons are $\bm{r}_i$ and $\bm{r}_{v_i}$, respectively.

The decisive approximation is now the choice of the Hilbert space allowed
for the wave function, that is
\begin{align}
\Psi= \mathcal{A} \left( \Phi_c(\bm{r}_1,\cdots,\bm{r}_A) \Phi_v(\bm{r}_{v_1},\bm{r}_{v_2}) \right),
\label{eq3}
\end{align}
where $\Phi_c = \det(\{\phi_i\})$, is the Slater determinant, of
single-particle wave functions, $\phi_i$, for the core nucleons,
$\Phi_v $ is the three-body wave function, and $\mathcal{A}$
symbolizes anti-symmetrization of all nucleons.  The form of $\Psi$ in
Eq.~(\ref{eq3}) clearly exhibits how we combine mean-field treatment
of the core and ordinary treatment of the two three-body relative
degrees of freedom.  The total energy, $E$, is a sum of two terms
corresponding to core, $E_c$, and valence, $E_v$, Hamiltonians in
Eq.~(\ref{hamcv}), that is
\begin{equation} \label{energycv}
 E = E_c + E_v = \langle \Psi|H_c|\Psi \rangle + 
 \langle \Psi|H_v|\Psi \rangle \; .
\end{equation}
We find the equations for the lowest energy solution by varying the
wave functions over the allowed Hilbert space, that is
\begin{equation} \label{varyeq}
\frac{\delta E}{\delta\Phi_c^*} = \frac{\delta E}{\delta\Phi_v^*} = 0\;,
\end{equation}
where both $\Phi_c$ and $\Phi_v$ must be normalized during the
variation.  The form of the two resulting equations are
\begin{equation} \label{schcoup}
  H_{HF}(\Phi_v,\Phi_c) \phi_i = \epsilon_i \phi_i \; 
; \hspace*{3mm}
  H_{v}(\Phi_v,\Phi_c) \Phi_v = E_v \Phi_v  \;,
\end{equation}
where the effective interactions in both Hartree-Fock single-particle,
$H_{HF}$, and three-body, $H_{v}$, Hamiltonians depend on both
$\Phi_c$ and $\Phi_v$.  The coupled equations in Eqs.~(\ref{schcoup})
must be solved simultaneously, which in practice means iteratively, to
determine $\Phi_c$ and $\Phi_v$, and subsequently the energy, in a
self-consistent procedure.

\paragraph{Interactions.}

We focus on the neutron dripline nucleus, $^{26}$O with the dominating
configuration of two neutrons around a $^{24}$O-core.  The choice of
mean-field approximation requires a corresponding effective
interaction. For the nucleon-nucleon interactions we choose the
popular Skyrme form \cite{vau72} with SLy4 parameters \cite{cha98}.
The three-body interaction in Eq.~(\ref{hamc}) is implemented as a
density dependent two-body interaction.

The Skyrme interaction is of zero range and therefore not directly
applicable in Hilbert spaces beyond Slater determinants.  The
necessary renormalization is possible but requiring additional
investigations. Instead we use the finite-range nucleon-nucleon
interaction in vacuum for $V_{A+1,A+2}$ in Eq.~(\ref{hamv})
\cite{gar99b}. We leave a more consistent adjustment to future
refinements, because this interaction has very little influence on the
small-distance structures, while more importantly the large-distance
asymptotic properties are correct \cite{gar99b}.

The phenomenological density dependence of the Skyrme interaction
parametrizes all otherwise omitted influences, e.g. three-body
effective forces.  These effects are all accounted for by the two-body
terms in Eqs.~(\ref{hamc}) and (\ref{hamv}) except $\sum_{i}^{A}
V_{i,A+1,A+2}$.  Again we leave a more consistent derivation to future
studies, because this term has very small structure influence, but it
is necessary, if fine tuning of the global Skyrme energy is needed.
We replace this term by $V_{c,A+1,A+2} = S_0 \exp(-\rho^2/\rho_0^2)$,
where the hyperradius $\rho$ is defined as
\begin{align}
(m_n+m_{c}) \rho^2 
&= m_{c} \left( (\bm{r}_{v_1} - \bm{R}_{c})^2 + (\bm{r}_{v_2} - \bm{R}_{c})^2 \right)
\notag 
\\
&+ m_n(\bm{r}_{v_1} - \bm{r}_{v_2})^2 \; ,
\label{rhodef}
\end{align}
where $m_n$, $m_{c}$ and $\bm{R}_{c}$ are neutron mass, core
mass and core center-of-mass coordinate, respectively.  The limit of
zero at large-distances is correct by construction.  The range and
strength parameters are $\rho_0=6$~fm and $S_0 = -6.45 $~MeV.  The
Skyrme interactions lead to density dependent effective masses, which
also appear in the coupled Eqs.~(\ref{schcoup}).  This is a new
feature in three-body equations and in few-body physics in general.

The space allowed for the valence nucleons is only limited by the
presence of the identical core nucleons. In the three-body calculation
these core-occupied Pauli forbidden states are removed either by
excluding the corresponding lowest adiabatic potentials or by
constructing phase equivalent potentials with less bound states
\cite{gar99}.  The voluminous and tedious, but straightforward,
derivation along with the subtleties and the space-requiring detailed
formulae will be discussed in forthcoming publications.

Meaningful combination and parallel treatment of core and valence
spaces require careful selection and perhaps adjustments of the
interactions.  Our choices are consistent but two less important links
between the interactions still needs to be fine-tuned.  Achieving
rigorous consistency is probably difficult in general and the most
obvious first application is on small- to large-distance dependence of
two-nucleon correlations around a finite nucleus.

\begin{figure}[t]
\centering
\includegraphics[width=1\linewidth]{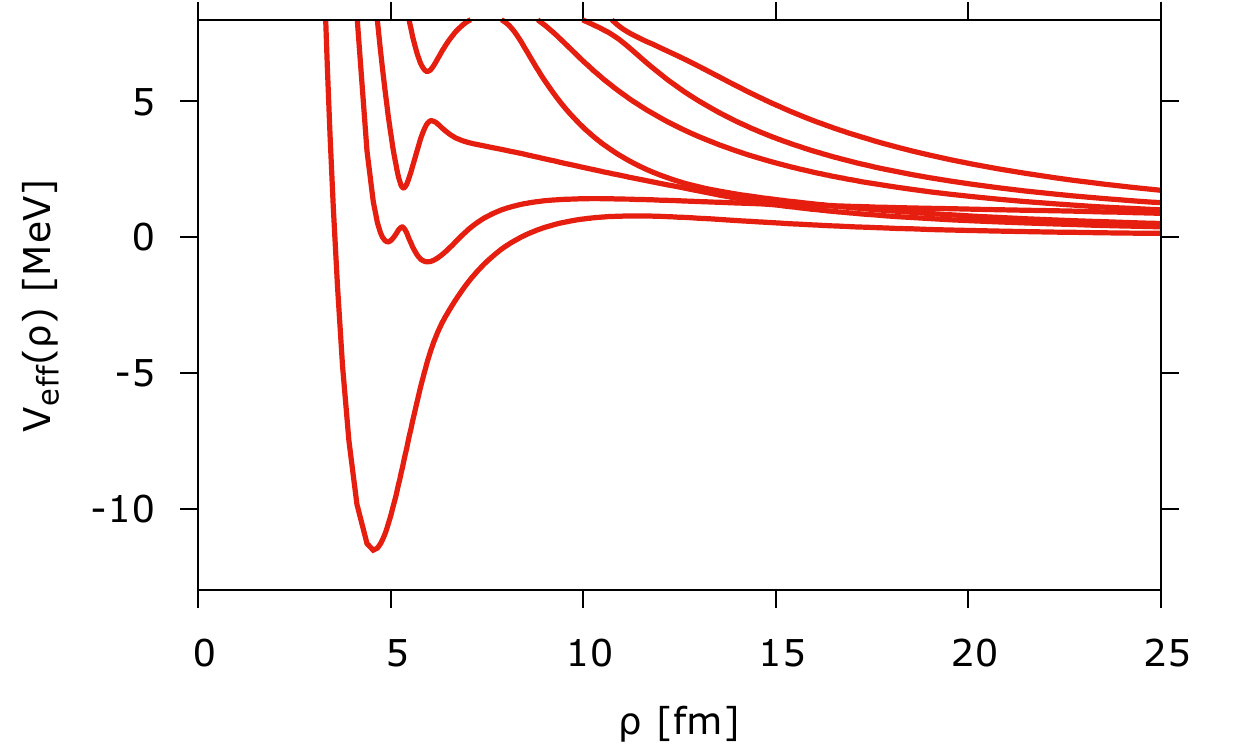}
\caption{The lowest adiabatic potentials for the SLy4 interaction,
  after removal of the Pauli forbidden states. The hyperradius,
  $\rho$, is normalized with the neutron mass \cite{nie01}. }
\label{fig1:lambda}
\end{figure}

\paragraph{Three-body energies.}

With these interactions we find from Eq.~(\ref{varyeq}) the
self-consistent variational solution where the core particles are
affected by the valence nucleons and vice versa.  We use the
hyperspherical adiabatic expansion method \cite{nie01} for the
three-body part where the basic ingredients are the adiabatic
potentials displayed in Fig.~\ref{fig1:lambda}.  The lowest potential
is attractive at small distances with a wide barrier of height $0.8$
MeV. The higher-lying potentials are mostly repulsive with features of
avoided crossings.

\begin{figure}
\centering
\includegraphics[width=1\linewidth]{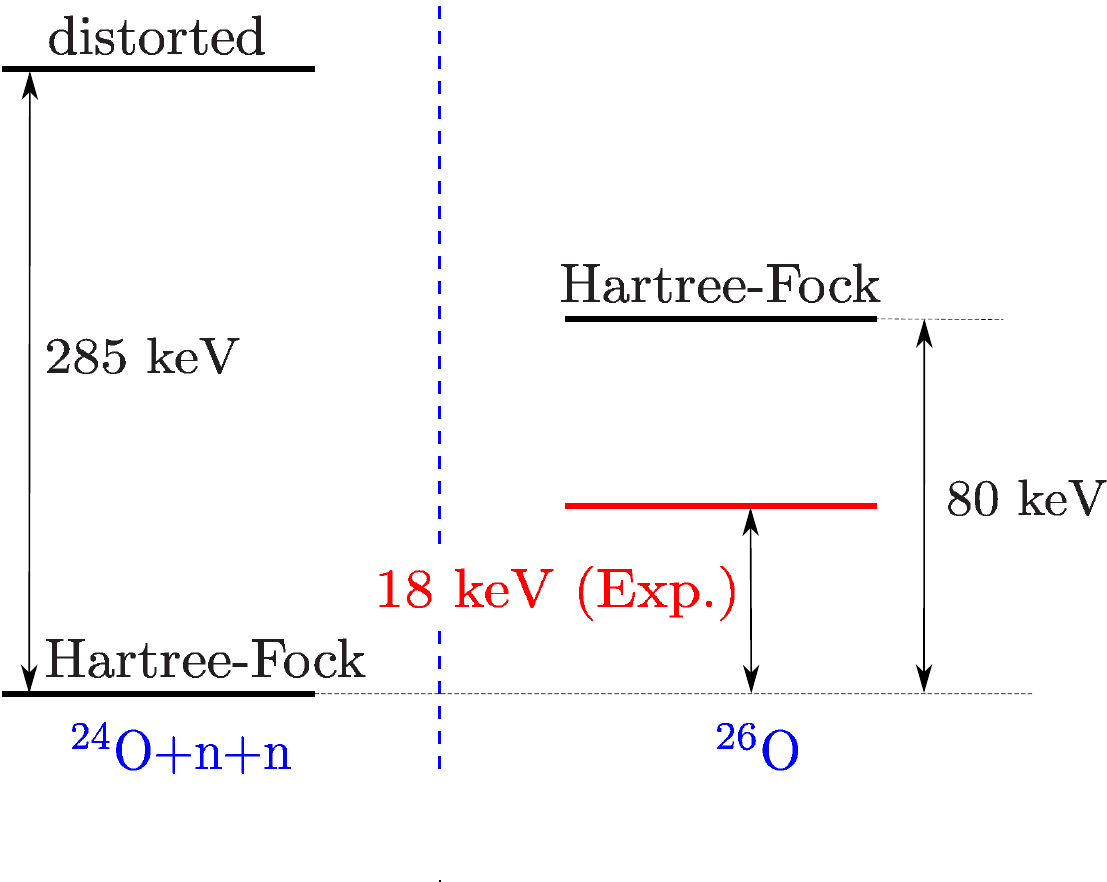}
\caption{The distortion energy of $^{24}$O (left) within the $^{26}$O
  system, and the ground state energies of $^{26}$O (right) for the
  Hartree-Fock approximation and the present method.  The zero-point
  is the energy of $^{24}$O in the Hartree-Fock ground state.  }
\label{ener}
\end{figure}

The \label{page:energy} three-body energy and wave function are found by solving the set
of coupled radial equations corresponding to these potentials
\cite{nie01}.  The energy, $E$($^{26}$O)$ = -172.490$~MeV, is obtained
for a wave function of the form in Eq.~(\ref{eq3}) which is $62$~keV
lower than obtained when the wave function is a pure Slater
determinant, $E_{HF}$($^{26}$O)$= -172.428$~MeV.  Both these energies
are higher than the energy, $E_{HF}$($^{24}$O)$= -172.508$~MeV,
obtained by moving the two neutrons infinitely far away, but
maintaining the wave function in Eq.~(\ref{eq3}).  Thus, the
two-neutron separation energy is $18$~keV in this picture.  The
influence of the two neutrons on the $^{24}$O-core is measured by its
distortion energy, which is 285 keV above $E_{HF}(^{24}\mbox{O})$,
within the ground state of $^{26}$O. Therefore, the two neutrons are bound
by $267$~keV with respect to the distorted $^{24}$O-nucleus.

These numbers are compared in Fig.~\ref{ener}, where we notice that
our method qualitatively has the correct properties.  The
$^{24}$O-nucleus is excited from its ground state by the presences of the two neutrons. The Hartree-Fock energy of $^{26}$O is higher than $E(^{26}\mbox{O})$,
which is reassuring since we should provide an improvement of the pure
mean-field calculation.  Apparently we have consistently extended both
interactions and Hilbert space beyond Slater determinants.  Thus, the
detailed structure of the wave function is an essential improvement in
our model.

\begin{figure}
\centering
\includegraphics[width=1\linewidth]{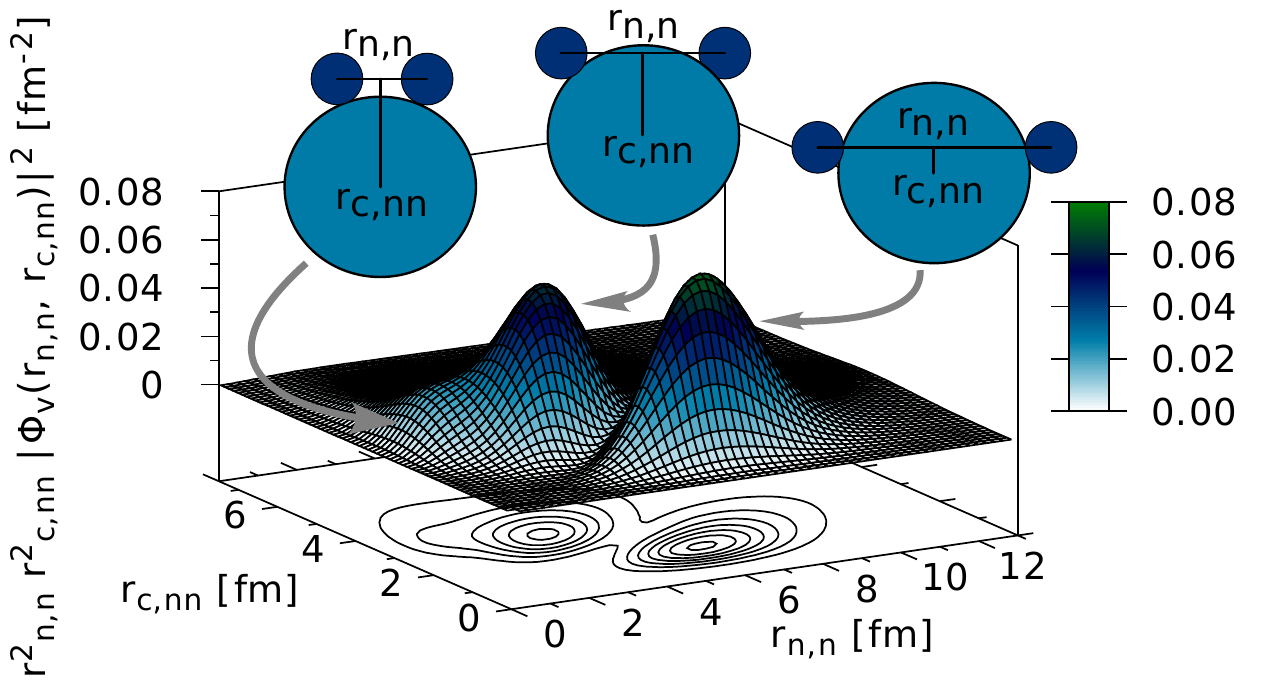}
\caption{The probability distribution of the two valence neutrons in
  $^{26}$O calculated using the Skyrme SLy4 interaction \cite{cha98}
  as a function of neutron-neutron distance, $r_{n,n}$, and core to
  neutron-neutron center of mass distance, $r_{c,nn}$. The insets are
  a schematic illustration of the configuration at the peaks. }
\label{probab}
\end{figure}

\paragraph{Three-body structure.}

The probability distribution is shown in Fig.~\ref{probab} as a
function of neutron-neutron distance, and core to neutron-neutron
center of mass distance. Two sharp peaks are seen corresponding to a
distance between the neutrons and their center of mass and the core of
about $(6,2)$~fm and $(3,3)$~fm, respectively. These are approximately
linear and equal sided triangular configurations as shown
schematically in Fig.~\ref{probab}. A much fainter peak at distances
(4,1.8)~fm is also seen temptingly interpreted as a di-neutron
signature. In Ref.~\cite{hag16} the same triple peak structure is
obtained using phenomenological interactions, but the di-neutron
configuration is concluded to be the dominating structure.

To compare properly we repeated the calculation using the neutron-core
potential given in \cite{hag16}, and found very similar peak
structures as in Fig.~\ref{probab}.  We find that the ``di-neutron''
configuration in Ref.~\cite{hag16} is much smaller than, but separated
from, the other two peaks.  The differences from us are due to
different adjustments of neutron-core Woods-Saxon potential to give
the $^{25}$O properties, and the density dependent neutron-neutron
pair potential to give the $^{26}$O energy with the use of the bare
nucleon mass. Especially the last adjustment differs from our model.

\begin{figure}
\centering
\includegraphics[width=1\linewidth]{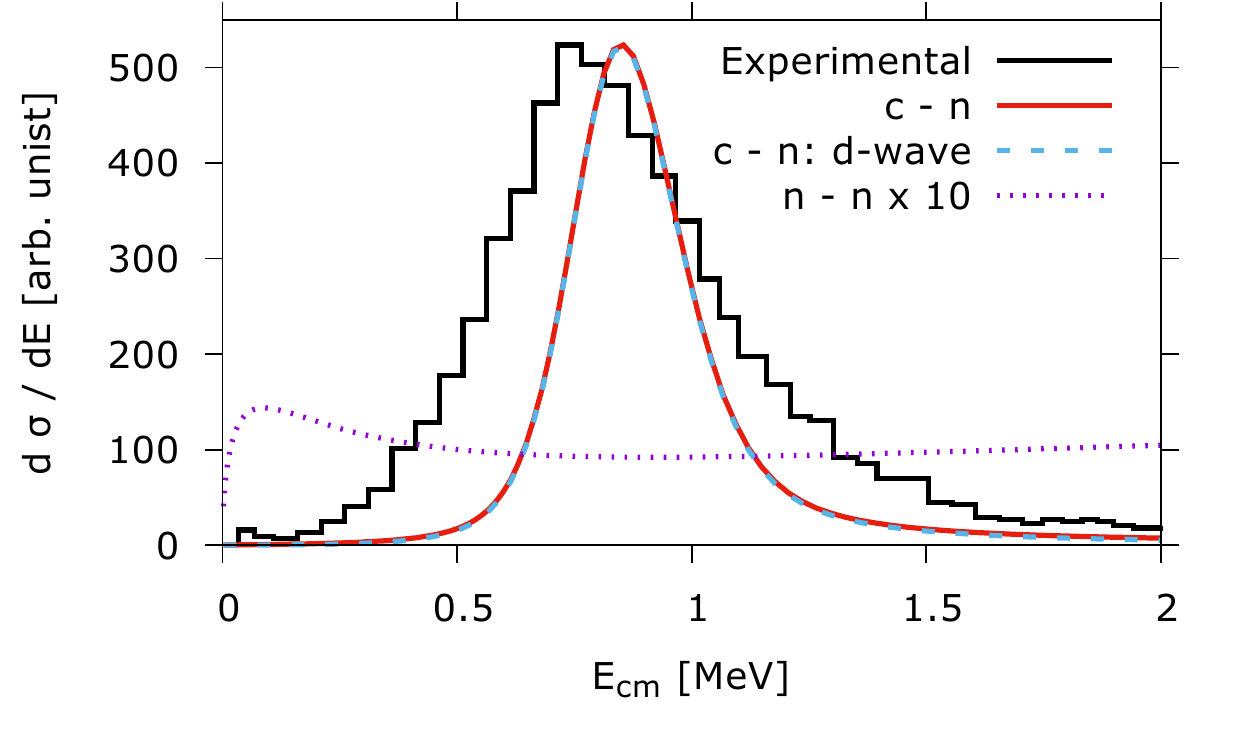}
\caption{The invariant mass spectra of core-neutron (solid) and neutron-neutron (dotted) for the SLy4 Skyrme parameters. The dashed curve is the core-neutron contribution from $d$-waves. The black curve is the measurements from Ref.~\cite{kon16}. }
\label{fig2:invarmass}
\end{figure}

We can also study the structure of the $^{25}$O ground state resonance
through the invariant mass spectra of two of the particles after
knockout of the third one \cite{zin97}.  The results are shown in
Fig.~\ref{fig2:invarmass}, where we first notice the expected
structureless neutron-neutron spectrum.  The neutron-core spectrum is
more interesting with a peak at $0.85$~MeV, which is only $0.1$~MeV higher
than the experimentally known $d_{3/2}$ resonance at $0.749(10)$~MeV
\cite{kon16}.

This is in fact a remarkably good agreement for three reasons.  First,
this result of the neutron-core resonance energy is obtained without
any free parameters, that is without any adjustment, and it is found directly
from the same interaction as between the nucleons in the core.  This is
also in contrast to phenomenological models \cite{hag16} where this
two-body energy is used as input parameter.  Second, a pure
Hartree-Fock calculation of $^{26}$O, with the same Skyrme force,
yields a $d_{3/2}$ energy of $-0.96$~MeV.  This is a bound state and
far from the experimental value, and as such revealing the inadequacy
of the Hartree-Fock approximation.  Third, the final state in $^{25}$O
is populated in two different reactions in experiment and theory,
i.e. by high-energy proton and neutron knockout, respectively.

The calculated width does not include effects of the unavailable
experimental resolution and therefore understandably smaller than the
observed value.  The neutron-core spectrum is almost indistinguishable
from the $d$-wave contribution also shown in
Fig.~\ref{fig2:invarmass}.  This reflects the structure of $90\%$
neutron-core $d_{3/2}$-wave in the total three-body wave function. The
rest is an equal distribution of $p_{3/2}$, $d_{5/2}$ and $f_{7/2}$
waves.

\begin{figure}
\centering
\includegraphics[width=1\linewidth]{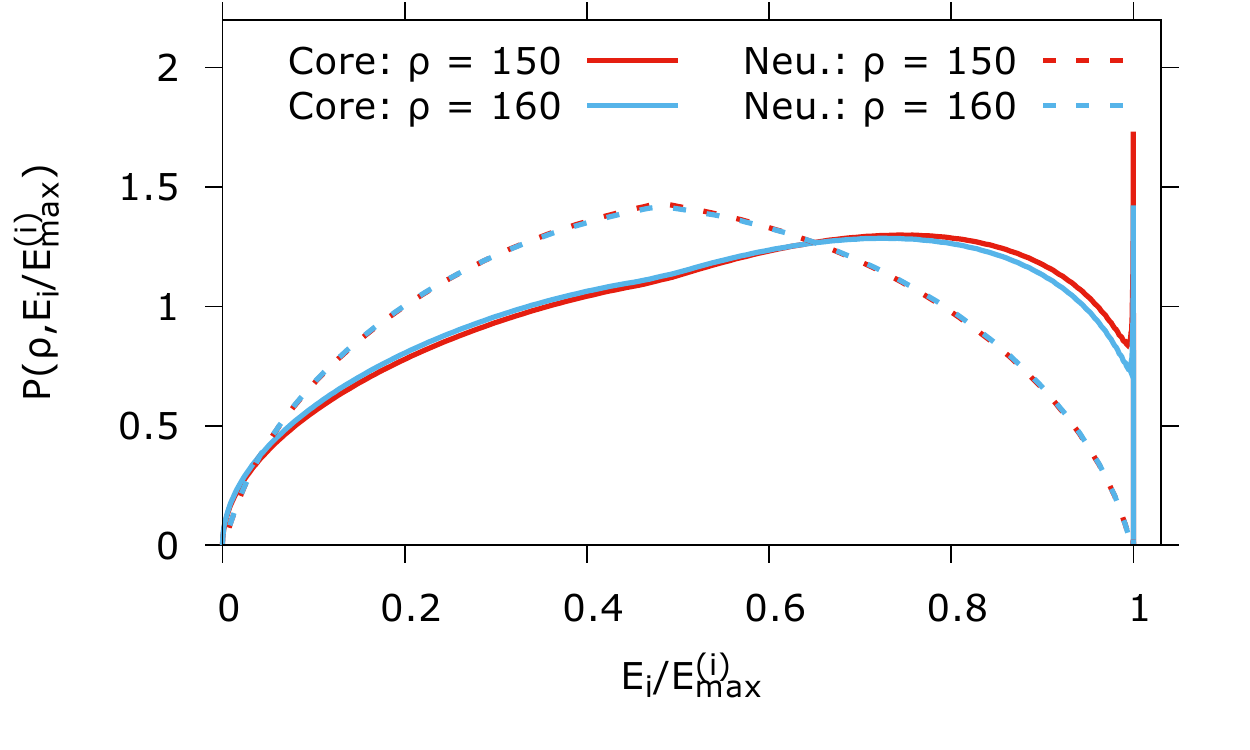}
\caption{ Single-particle energy distributions after decay of the
  ground state resonance \cite{gar07,fyn09}.  Here $E_i$ is core
  (solid) or neutron (dashed) energy, and $E^{(i)}_{max}$ is the
  maximum energy available for particle $i$. The peak for the core
  energy is due to numerical inaccuracies.}
  \label{energydistrib}
\end{figure}

\paragraph{Lifetime and decay properties}

As already mentioned, a central aspect of the present method is how
the presence of the valence nucleons affects the core. This is
reflected in distortion of the core wave function and therefore also
in the energy. By going from small to large values of hyperradius,
$\rho$, the core should change from a distorted structure to the free
solution.  However, currently only the average effect is included and
the core structure is maintained for all distances.  This lack of a
gradual relaxation means that while the bulk part of the calculated
potential in Fig.~\ref{fig1:lambda} is correct, the large-distance
asymptotic value of the potential is too high, in this case by
$285$~keV.  However, by extending the adiabatic approach, and solving
the coupled expressions from Eq.~(\ref{schcoup}) for each step in
$\rho$, a smooth transition could be obtained.  We leave this more
elaborate procedure for future improvements, because this would only
marginally change any of the observables except possibly the width.

This width of the $^{26}$O ground state resonance can still be fairly
well estimated in the present implementation as the
oscillator-approximated knocking rate multiplied by the WKB tunneling
probability through the modified barrier in the lowest adiabatic
potential \cite{gar04b}. First we shift the full adiabatic potential
by 285 keV.  This leads to an outer turning point of $66.3$~fm and a
lifetime of $10^{-16}$~s.  Instead, when only lowering the potential
by $285$~keV in the tail outside $66.3$~fm, the well-defined outer
turning point at the energy of $18-285=-267$~keV leads to a lifetime
of about $10^{-15}$~s.  Experimentally the half-life is found to be
between $10^{-17}$~s and $10^{-15}$~s \cite{kon16}, and we conclude
that even when employing a very crude transition mechanism this
exponentially sensitive observable is predicted remarkably well by the
model.

The resonance is decaying into two neutrons and the core. Following
\cite{gar07,fyn09}, the single-particle energy distributions after
decay are obtained in coordinate space from the stable spatial
distribution of the particles for large values of $\rho$, where the
hyperangles in coordinate and momentum space coincide. The results are
shown in Fig.~\ref{energydistrib} for two $\rho$-values, $150$ and
$160$~fm, to show convergence.  The extremes, where one particle takes
either maximum or zero energy, leaves either zero or maximum energy in
the relative motion between the other two particles.

We then see in Fig.~\ref{energydistrib} that the core energy is
weighted towards its maximum with a corresponding decreasing fraction
left for the relative motion of the neutrons (solid curves).  Each
neutron has largest probability for appearing with half of its maximum
energy, again implying that the other half is in relative neutron-core
motion (dashed curves).  This shows that the decay mechanism is direct
population of the continuum, consistent with the fact that the
$d_{3/2}$ resonance in $^{25}$O is too high in energy to be even
virtually populated during the decay.

\paragraph{Summary.}  

The present study provides a consistent approach to including the
intricacies of few-body formalisms into a many-body context.  The
model is practical and efficient as demonstrated in the application on
the challenging nuclear neutron dripline nucleus, $^{26}$O.  The data
of both $^{25}$O and $^{26}$O are reproduced with fewer parameters
than found in dedicated phenomenological models. Furthermore, the
exponentially sensitive lifetime is obtained within measured
uncertainties.  The novel features of the quantum mechanical model are
that few- and many-body properties respectively at large and small
distances are self-consistently connected.  The model is applicable to
both bound and continuum states, and addresses challenges in open
quantum systems.  The universal character of halos and Efimov states
suggests applications in other subfields of physics.

\paragraph{Acknowledgment}

This work was funded by the Danish Council for Independent Research
DFF Natural Science and the DFF Sapere Aude program. This work has been partially supported by the Spanish Ministerio de Economia y Competitividad under Project FIS2014-51971-P. We acknowledge many fruitful discussions with Karsten Riisager.

\end{document}